\documentclass{ws-procs975x65}
\newcommand{\be}{\begin{equation}}
\newcommand{\ee}{\end{equation}}
\newcommand{\bea}{\begin{eqnarray}}
\newcommand{\eea}{\end{eqnarray}}
\begin{document}


\title{The significance of matter coupling in $f(R)$ gravity}

\author{Thomas P.~Sotiriou\footnote{sotiriou@sissa.it}}
\address{SISSA-International School for Advanced Studies, via Beirut 2-4, 34014, Trieste, Italy and\\INFN, Sezione di Trieste}


\begin{abstract}
The way one chooses to couple gravity to matter is an essential  
characteristic
of any gravitational theory. In theories where the gravitational  
field is
allowed to have more degrees of freedom than those of General Relativity
(\emph{e.g.~}scalar-tensor theory, $f(R)$ gravity)
this issue often becomes even more important.
We  
concentrate here on $f(R)$ gravity treated within the
Palatini variational principle and discuss how the coupling  
between matter and
the extra degrees of freedom of gravity (the independent  
connections in our
case) affects not only the resulting phenomenology but even the geometrical meaning of fundamental fields.\\
\end{abstract}

\bodymatter
Some of the most prominent questions in Physics are nowadays related to gravity. These questions are relevant for High Energy Physics, since finding a quantum theory of gravity has proved to be a difficult task and theories currently considered for the description of gravity at small scales seems to imply that non trivial corrections with respect to the Einstein-Hilbert term should be included in the gravitational action \cite{Buch}. They are also relevant for Cosmology and Astrophysics, 
since 
current 
observations indicate that approximately 70\% of the energy density of the universe is due to an unkown form of energy, which is often called dark energy and is considered to be responsible for the late time accelerated expansion of the universe\cite{Spergel:2006hy}. It is reasonable to examine the possibility that these problems are actually related: corrections in the effective low energy gravitational action coming from our high energy theories can lead to modified gravity even at large scales, which might account for dark energy.

One of the easiest modifications of gravity comes from generalizing the Einstein-Hilbert action by assuming that the gravitational Lagrangian is a general function of the scalar curvature, $f(R)$. But is $f(R)$ gravity really the answer? Possibly not, but this is actually the wrong question to ask! Since we still have little evidence about which corrections to the gravitational action are most likely to appear --- this is theory dependent and there are numerous candidates for ``quantum'' gravity --- we seem to be more in need of a toy low energy gravitational theory to understand how different corrections influence our picture for the gravitational interaction. And $f(R)$ gravity seems to be very suitable for these purposes.

If $S_M$ is used to denote the matter action, the action for $f(R)$ gravity is 
\be
\label{action}
S=\frac{1}{16 \pi G}\int d^4 x \sqrt{-g} f(R)+S_M.
\ee
For  $f(R)=R$ it reduces to the Einstein-Hilbert action and metric 
variation yields the Einstein equations. As can be found in textbooks\cite{textbooks}, an alternative is to use the Palatini variation, {\em i.e.~}an independent variation of the metric and connection which gives the same field equations but also the expression for the connection. For doing this we take $R=g^{\mu\nu}R_{\mu\nu}(\Gamma)$, where $R_{\mu\nu}(\Gamma)$ is the Ricci tensor constructed with the independent connection $\Gamma^\lambda_{\phantom{a}\mu\nu}$. What needs to be stressed here is that in order to derive the Einstein equations with the Palatini variation one has to make  an extra assumption: that $S_M$ does not depend of the connection, and  hence
\be
\label{assumption}
\frac{\delta {S}_M}{\delta \Gamma^\lambda_{\phantom{a}\mu\nu}}=0.
\ee
Under this assumption, the Palatini variation yields the field equations\cite{vol}
\bea
\label{f1}
f'(R) R_{(\mu\nu)}-\frac{1}{2}f(R)g_{\mu\nu}&=&8 \pi G T_{\mu\nu},\\
\label{f2}
\nabla_\lambda \left( \sqrt{-g}f'(R)g^{\mu\nu}\right)&=&0,
\eea
which for $f(R)=R$ reduce to Einstein's equations and the definition of the Levi-Civita connection respectively. The Palatini variation together with the assumption stated in eq.~(\ref{assumption}) constitute was is called the Palatini formalism or Palatini $f(R)$ gravity, if the action has the form of (\ref{action}). For a more general choice of $f$, eq.~(\ref{f2}) implies that $\Gamma^\lambda_{\phantom{a}\mu\nu}$ is the Levi-Civita connection of the metric $h_{\mu\nu}=f'(R)g_{\mu\nu}$, and therefore eq.~(\ref{f1}) can be written as an equation for $g_{\mu\nu}$, which, however, is different from both Einstein's equations and the field equations of $f(R)$ gravity that one derives by standard metric variation \cite{buchdahl}.

What I want to focus on here is that the assumption (\ref{assumption}) is physically meaningful and not trivially satisfied. The matter Lagrangians for scalar fields or the electromagnetic field do not depend on the connection. However, this is not true for all matter fields, for example fermions. Therefore, forcing this assumption can mean only two things: either only certain matter fields are included in the theory, or, for some reason, Dirac fields or other matter fields that generally couple to the connection, couple to the Levi-Civita connection of the metric instead of the independent one. The first option does not seem suitable for a theory describing the gravitational interaction, since such a theory would be very limited. The second option is not very appealing either. If the independent connection $\Gamma^\lambda_{\phantom{a}\mu\nu}$ is to have the usual geometrical properties, such as defining parallel transport and the covariant derivative, then indeed this is the connection that Dirac fields should be coupled to. Actually in this case, it is even more appropriate if this connection is allowed to be non-symmetric. The resulting theory is then a metric-affine theory of gravity\cite{sotlib}, whose field equations resemble eqs.~(\ref{f1}) and (\ref{f2}) but torsion terms and more importantly the matter term $\Delta_{\lambda}^{\phantom{a}\mu\nu}\equiv-\frac{2}{\sqrt{-g}}\frac{\partial {\cal S}_M}{\partial \Gamma^\lambda_{\phantom{a}\mu\nu}}$ are present in the equivalent of eq.~(\ref{f2}).


The fact, however, that assuming {\it a priori} that matter fields are coupled to the Levi-Civita connection instead of the independent one is not very appealing, does not mean that it is unfeasible. On the other hand, it does raise questions about the geometrical meaning of $\Gamma^\lambda_{\phantom{a}\mu\nu}$ which is obviously not related with parallel transport or the covariant derivative (unlike in metric-affine gravity). One can also show that in the Palatini formalism it is not even related to matter conservation laws, since these are expressed using the covariant derivate related to the metric\cite{Koivisto:2005yk}. Actually,  if $\Gamma^\lambda_{\phantom{a}\mu\nu}$ does not couple to matter, then the action (\ref{action}) is dynamically equivalent to the action of scalar-tensor theory with Brans-Dicke parameter $\omega_0=-3/2$ \cite{Sotiriou:2006hs}. Notice, that this means that there is only one extra scalar degree of freedom besides the metric, even thought $\Gamma^\lambda_{\phantom{a}\mu\nu}$ has $64$ component, $40$ of which are initially independent if it is assumed to be symmetric! Therefore, applying the Palatini variation whilst assuming (\ref{assumption}), is different from assuming that the metric and the true connection of spacetime are independent. The true connection in this case is the Levi-Civita one {\it a priori} and $\Gamma^\lambda_{\phantom{a}\mu\nu}$ is an auxiliary field whose introduction serves only to add a scalar degree of freedom contrary to what one might think by examining the action.


Summarizing, it is worth mentioning the following: Claiming that the Palatini variation leads to the Einstein equation directly is imprecise. Even when the Einstein-Hilbert action is used, one needs the extra assumption that the independent connection should not be coupled to the matter. The physical meaning of this assumption is that this connection does not define the covariant derivative and is therefore not the true connection of spacetime but an auxiliary field void of geometrical meaning. The above can also be seen through the equivalence of Palatini $f(R)$ gravity with scalar tensor theory. If no such assumption is made then one gets a metric-affine theory of gravity, which will have the same phenomenology as Palatini $f(R)$ gravity in cases where only matter fields that naturally do not couple to the connection are considered, such as in cosmology. 

{\em Acknowledgements:} The author wishes to thank Stefano Liberati and John Miller for fruitful discussions and valuable comments.

\vfill

\end{document}